\documentclass[11pt,twoside]{article}
\usepackage{amsfonts,amssymb}
\usepackage{amsmath}
\def\sqr#1#2{{\vcenter{\hrule height.#2pt\hbox{\vrule width.#2pt
height#1pt \kern#1pt \vrule width.#2pt}\hrule height.#2pt}}}
\def\square{\mathchoice\sqr64\sqr64\sqr{4.2}3\sqr{3.0}3 \ \!\!\!}

\def\hook{\hbox{\vrule height0pt width4pt depth0.3pt
\vrule height7pt width0.3pt depth0.3pt \vrule height0pt width2pt
depth0pt} }
\newcommand{\sect}[1]{\setcounter{equation}{0}\bigskip\medskip
\section{#1}\smallskip}

\newtheorem{THEOREM}{Theorem}[section]  
\newenvironment{theorem}{\begin{THEOREM} \hspace{-.85em} {\bf :} 
}%
                        {\end{THEOREM}}
\newtheorem{LEMMA}[THEOREM]{Lemma}
\newenvironment{lemma}{\begin{LEMMA} \hspace{-.85em} {\bf :} }%
                      {\end{LEMMA}}
\newtheorem{COROLLARY}[THEOREM]{Corollary}
\newenvironment{corollary}{\begin{COROLLARY} \hspace{-.85em} {\bf 
:} }%
                          {\end{COROLLARY}}
\newtheorem{PROPOSITION}[THEOREM]{Proposition}
\newenvironment{proposition}{\begin{PROPOSITION} \hspace{-.85em} 
{\bf :} }%
                            {\end{PROPOSITION}}
\newtheorem{DEFINITION}[THEOREM]{Definition}
\newenvironment{definition}{\begin{DEFINITION} \hspace{-.85em} {\bf 
:} \rm}%
                            {\end{DEFINITION}}
\newtheorem{EXAMPLE}[THEOREM]{Example}
\newenvironment{example}{\begin{EXAMPLE} \hspace{-.85em} {\bf :} 
\rm}%
                            {\end{EXAMPLE}}
\newtheorem{CONJECTURE}[THEOREM]{Conjecture}
\newenvironment{conjecture}{\begin{CONJECTURE} \hspace{-.85em} 
{\bf :} \rm}%
                            {\end{CONJECTURE}}
\newtheorem{PROBLEM}[THEOREM]{Problem}
\newenvironment{problem}{\begin{PROBLEM} \hspace{-.85em} {\bf :} 
\rm}%
                            {\end{PROBLEM}}
\newtheorem{REMARK}[THEOREM]{Remark}
\newenvironment{remark}{\begin{REMARK} \hspace{-.85em} {\bf :} 
\rm}%
                            {\end{REMARK}}
\newtheorem{CONCLUSION}[THEOREM]{Conclusion}
\newenvironment{conclusion}{\begin{CONCLUSION} \hspace{-.85em} {\bf :} 
\rm}%
                            {\end{CONCLUSION}}
\newcommand{\thm}{\begin{theorem}}
\newcommand{\lem}{\begin{lemma}}
\newcommand{\pro}{\begin{proposition}}
\newcommand{\dfn}{\begin{definition}}
\newcommand{\rem}{\begin{remark}}
\newcommand{\con}{\begin{conclusion}}
\newcommand{\xam}{\begin{example}}
\newcommand{\cnj}{\begin{conjecture}}
\newcommand{\prb}{\begin{problem}}
\newcommand{\cor}{\begin{corollary}}

\newcommand{\ethm}{\end{theorem}}
\newcommand{\elem}{\end{lemma}}
\newcommand{\epro}{\end{proposition}}
\newcommand{\edfn}{\bbox\end{definition}}
\newcommand{\erem}{\bbox\end{remark}}
\newcommand{\econ}{\bbox\end{conclusion}}
\newcommand{\exam}{\bbox\end{example}}
\newcommand{\ecnj}{\bbox\end{conjecture}}
\newcommand{\eprb}{\bbox\end{problem}}
\newcommand{\ecor}{\end{corollary}}

\newcommand{\beqn}{\begin{equation}}
\newcommand{\eeqn}{\end{equation}}

\newcommand{\bbox}{\vrule height7pt width4pt depth1pt}

\def\br{\begin{eqnarray}}
\def\er{\end{eqnarray}}
\def\brn{\begin{eqnarray*}}
\def\ern{\end{eqnarray*}}
\def\er{\end{eqnarray}}
\def\eqq{&\!\!\!\!\!=\!\!\!\!\!&}

\def\vt{\vartheta}

\def\L{{\cal{L}}}

\def\a{\alpha}
\def\b{\beta}
\def\g{\gamma}
\def\d{\delta}
\def\g{\gamma}

\def\L{\mathcal{L}}

\title{{\bf On a class of invariant coframe operators with application 
to gravity}\\
}
\author{
\thanks {\quad   email itin@math.huji.ac.il}
\small {\rm Yakov Itin and 
\thanks {\quad   email kaniel@math.huji.ac.il}
Shmuel Kaniel}\\
\small {\rm Institute of Mathematics}\\
\small {\rm Hebrew University of Jerusalem}\\
\small {\rm Givat Ram, Jerusalem 91904, Israel}\\
}
\begin{document}  

\pagestyle{myheadings}
\markboth{Y. Itin \&  S. Kaniel \qquad}{On a class of invariant coframe operators with application to gravity}
\newcommand{\bi}[1]{\bibitem{#1}}
\date{}

\maketitle
\begin{abstract}
\rm{Let a differential $4D$-manifold with a smooth coframe field be given. 
Consider the operators on it that are linear in the second order 
derivatives or quadratic in the first order derivatives of the coframe, 
both with  
coefficients that depend on the coframe variables.
The paper exhibits the class of operators that are invariant under 
a general change of coordinates, and, also, invariant under the global $SO(1,3)$-transformation of the coframe. A general  class of field equations is constructed.   
We display two subclasses in it. 
The subclass of field equations that are derivable from action principles by free variations and the subclass of field equations  for 
which spherical-symmetric solutions, Minkowskian at infinity exist. 
Then, for the  spherical-symmetric solutions, the resulting metric is computed. 
Invoking the Geodesic Postulate, we find all the equations  that are experimentally 
(by the 3 classical tests) indistinguishable  from Einstein field equations. 
This family includes, of course, also  Einstein equations. Moreover, it is shown, explicitly, how to exhibit it. 
The basic tool employed in the paper is an invariant formulation reminiscent 
of Cartan's structural equations. 
The article sheds light on the possibilities and limitations of the coframe gravity. It may also 
serve as a general procedure to derive covariant field equations. }\\
\it {PACS: 02.40.Hw, 04.20.Cv }
\end{abstract}
\sect{Introduction}
\rm{
In the framework of the Einstein theory the gravity field is described as geometrical property of a four-dimensional pseudo-Riemannian manifold (differential manifold endowed with a pseudo-Euclidean metric). In this manifold the evolution of the metric is described by a field equation which is covariant i.e. invariant under the group of diffeomorphic transformations of the manifold. The metric is used as the basic building block. Thus, component-wise, the field equation is a system of 10 differential equations for 10 components of the metric tensor. Furthermore, Einstein looked for the second order PDE's, linear in the principle part. It is interesting to note that Einstein in General Relativity as well as his later writings on unified field theories did not rely the Action Principle.\\
After the establishment of General Relativity E. Cartan \cite{Car} introduced the notion of a frame to differential geometry. He, then, showed that the basic constructs of classical differential geometry can be obtained via his Repere Mobile.
Einstein   \cite{E1}, \cite{EM} applied Cartan's 
ideas for the definition of a teleparallel space. He attempted, by that, to construct a unified theory  of gravity and electromagnetism. 
 Weitzenb\"{o}ck \cite{We1} investigated of the geometric 
structure of teleparallel spaces.
Theories based on this geometrical structure are used for alternative models 
of gravity and also to describe the spin properties of matter.
For the recent investigations 
in this area see Refs. \cite{Tr} to\cite{M-H},  
and \cite {Kawai1} to   \cite {N-Y}.\\
For an account of teleparallel spaces in the metric-affine 
framework see \cite{hehl95}, \cite{G-H} .\\
In \cite{K-I} an alternative gravity model based on teleparallel spaces was suggested. Let us exhibit here a short account of it. There, the field equation was taken to be 
\begin{equation}
\label{1-1}
\square  \ \vt^a=\lambda\vt^a,
\end{equation}
where $\square \ =d*d*+*d*d$ is the Hodge - de Rham Laplacian.
It turns out that there exists a unique (up to a constant, which is 
identified with the mass) spherical-symmetric, asymptomatically flat and static solution:
\begin{equation}
\label{1-2}
\vt^0=e^{-\frac mr}dt \qquad \vt^i=e^{\frac mr}dt \qquad i=1,2,3,
\end{equation}
The resulting line element is
\begin{equation}
\label{1-3}
ds^2=e^{-2\frac mr}dt^2-e^{2\frac mr}(dx^2+dy^2+dz^2).
\end{equation}
The metric defined by (\ref{1-3}) is the celebrated Yilmaz-Rosen metric 
\cite{yilmaz58},\cite{yilmaz76}),\cite{rose73}, \cite{rose74}.  
This metric is experimentally (by three classical tests) indistinguishable from the Schwarzschild metric.
In \cite{K-I} it was shown that (\ref{1-1}) is derivable from a constrained 
variational principle. U. Muench, F. Gronwald and  F.\,W.~Hehl \cite{Hehl} 
placed the model in the area of the various teleparallel theories. They also 
demonstrated that the equation (\ref{1-1}) can not be derived from a 
variational principle, by  unconstrained variations. \\
In the present paper we study the structure of the invariant differential coframe operators on a teleparallel manifold. We construct a general field equation on the coframe variable. That is covariant and $SO(1,3)$ invariant. It is a system of 16 PDE's for 16 coframe variables. The equations are linear in the second order derivatives and quadratic in the first derivatives.
It turns out  that in contrast to the metric gravity there exists on a  teleparallel space  a wide class of invariant field equation.  \\
We also study the structure of field equations that  can be derived   from a quadratic Lagrangian by free variations. We show that these equations form a subclass of the class above. We also show that the Einstein field equation is a unique symmetric equation in this subclass.\\
Another subclass of equations is defined by the conformity to observations i.e confirmation  with the three classical tests of gravity. \\
The aproach can also be useful for the establishment of alternative field equations for other fields.\\
All the computations are carried out, explicitly, in a covariant manner. This introduces a great computational simplicity.

\sect{Invariant objects on the teleparallel space}
Consider a $4D$-differential manifold $M$ endowed with a smooth coframe tetrad  $\vt^a, \ a=0,1,2,3$. This is a basis of the cotangent space 
$\Lambda^1:=T^*_xM$ at an arbitrary  point $x\in M$.  
Let the vector space $\Lambda^1$ 
be endowed with the Lorentzian metric $\eta_{ab}=\eta^{ab}=diag(1,-1,-1,-1)$. 
We will refer to the triad $\{M,\vt^a,\eta_{ab}\}$ as a \textsl{teleparallel space}. \rm A hyperbolic metric on the manifold $M$ is defined by the 
coframe $\{\vt^a(x)\}$ as
\begin{equation}
\label{2-1}
g=\eta_{ab}\vt^a\otimes\vt^b.
\end{equation}
The coframe $\{\vt^a(x)\}$ is pseudo-orthonormal with respect  to the metric $g$  
$$g^{\mu\nu}\vt^a_\mu\vt^b_\nu=\eta^{ab}.$$ 
Let us exhibit our basic construction.\\
Consider the exterior differential of the basis 1-forms
\footnote{We use here and later  abbreviation 
$\vt^{ab\cdots}=\vt^a\wedge \vt^b\wedge \cdots$.}
\begin{equation}\label{2-3}
d\vt^a=\vt^a_{\b,\a}dx^\a\wedge dx^\b=\frac 12 {C^a}_{bc}\vt^{bc}.
\end{equation}
Where, for uniqueness, the coefficients ${C^a}_{bc}$ are  antisymmetric: 
$${C^a}_{bc}=-{C^a}_{cb}.$$ 
We will refer to the coefficients ${C^a}_{bc}$ as to 3-indexed $C$-objects.\\ 
These coefficients can be written explicitly as
\begin{equation}\label{2-3a}
{C^a}_{bc}=e_c\hook(e_b\hook d\vt^a).
\end{equation}
Their contraction i.e. 
$$C_a={C^m}_{bm}=-{C^m}_{mb}$$ 
will be referred to as the 1-indexed $C$-object.\\
Their explicit expression is 
\begin{equation}\label{2-3b}
C_a=e_m\hook(e_a\hook d\vt^m).
\end{equation}
These $C$-objects represent diffeomorphic covariant and $SO(1,3)$ invariant 
generalized derivatives of the coframe field $\vt^a$. In \cite{it2} it is shown that the coderivative of the coframe can also be represented by the $C$-objects. \\
In order to construct the invariant generalized  second order derivative we 
consider the exterior differential of the $C$-objects.  
\begin{equation}\label{2-4}
d{C^a}_{mn}={B^a}_{mnp}\vt^p.
\end{equation}
The components of this 1-form ${B^a}_{mnp}$ will be referred 
to as 4-indexed $B$-objects. 
Again, the ${B^a}_{mnp}$ are scalars. Observe that they are antisymmetric 
in the middle indices 
$${B^a}_{mnp}=-{B^a}_{nmp}.$$ 
The explicit expression of 4-indexed $B$-object is
\begin{equation}\label{2-4a}
{B^a}_{bcd}=e_d\hook\Big(e_c\hook(e_b\hook d\vt^a)\Big).
\end{equation}
As we will see later the field equation should include only the 2-indexed values. These can be obtained by a contraction of 4-indexed $B$-objects with 
$\eta_{ab}$. Because of the antisymmetry the only possible contractions are 
\begin{eqnarray}
\label{2-5}
{}^{(1)}B_{ab}&=&{B_{abm}}^m,\\
\label{2-6}
{}^{(2)}B_{ab}&=&{B^m}_{mab},\\
\label{2-7}
{}^{(3)}B_{ab}&=&{B^m}_{abm},
\end{eqnarray}
where the indices are raised and lowered via the $\eta_{ab}$, 
for example, ${B_{abm}}^m=\eta^{mk}\eta_{ap}{B^p}_{bmk}$.\\
This contracted objects will be  referred to as 2-indexed $B$-objects. 
Observe that one of them, namely${}^{(3)}B_{ab}$, is antisymmetric and two others have generally non-zero symmetric and antisymmetric parts. \\
Note, also, that the exterior differential of the 1-indexed $C$-object can be 
expressed by the 2-indexed $B$-object 
\begin{equation}\label{2-8}
dC_a=d{C^m}_{am}={B^m}_{amp}\vt^p=-{}^{(2)}B_{ap}\vt^p.
\end{equation}
We will also include in the general equation the scalar (non-indexed) objects constructed from the ${B^a}_{nmp}$.
By the antisymmetry of the middle indices only one non trivial 
full contraction $B$ 
of the quantities ${B^a}_{nmp}$ is possible (up to a sign):
\begin{equation}\label{2-9}
B={{B^a}_{ab}}^b= {B^a}_{abc}\eta^{bc}={}^{(1)}{B^a}_a={}^{(2)}{B^a}_a.
\end{equation}
The fully contracted $B$-object with no index attached will be referred to as the scalar 
 $B$-object.\\
The coframe $\vt^a$ is usually used as a device to express physical variables like metric and connections. We take the view that the $\vt^a$ may serve as physical variables, as well. 
As such we construct field equations for it. Let us list the conditions that we are impose on these equations. 
\begin{itemize}
\item The field equation should be global $SO(1,3)$-invariant and
tensorial diffeomorphic covariant. For that the $B$ and $C$-objects may served as building blocks.  
Thus we construct it from the $B$ and $C$-objects.
\item The coframe field $\vt^a$ has 16 independent components, these components
are the independent dynamical variables.  Thus, if we write the equation in a scalar form, it should be a 2-indexed equation. 
\item We are interested only in partial differential equations of the second order. Moreover,   we are looking for equations that are linear in the $B$-objects (so that an approximation by the wave operators is possible). 
\item The nonlinear part of field equation is taken to be quadratic in the $C$-objects. Thus, if the coframe field $\vt^a$ is dimensionless, all the parts of the equation have the same  dimension and all the free parameters are dimensionless.
\end{itemize}  
In the sequel we will develop a general procedure for getting  all the field equations that are constructed by the $C$ and $B$ objects. It is a wide class. In particular we get all the equations derived by the variation of a general quadratic Lagrangian. The Einstein equation is a particular case of our general field equation. The procedure above allows us to treat also the situation where the metric tensor $g$, as defined by (\ref{2-1}), is the primary physical variable. In this case there are 10 independent variables which are the combinations of the components of the coframe  $\vt^a$. These variables satisfy 10 field equations that turn out to be the Einstein equations. This will be worked out explicitly in the sequel. \\   
Let us write the leading (second order) part of the equation 
as a linear combination of the two-indexed  
$B$-objects:
\begin{equation}\label{2-12}
L_{ab}=\b_1 \ {}^{(1)}\!B_{(ab)}+\b_2 \ {}^{(2)}\!B_{(ab)}+
\b_3 \ {}^{(3)}\!B_{ab}+
\b_4 \ \eta_{ab}B+\b_5 \ {}^{(1)}\!B_{[ab]}+\b_6 \ {}^{(2)}\!B_{[ab]},
\end{equation}
where the symbols $(ab)$ and $[ab]$ mean, consequently, symmetrization and 
antisymmetrization of the indices. The coefficients $\b_i$ are free numerical constants.\\
(\ref{2-12}) is the general 2-indexed tensorial expression  constructed by 
 the 4-indexed object ${B^a}_{bcd}$ by combination of contraction and transpose operators. 
The scalar $B$ is transformed into a two indexed object by multiplying 
it with $\eta_{ab}$. Note that the terms in (\ref{2-12}) are not independent. Their number will be reduced. \\
The general quadratic part of the equation can be constructed as 
a linear combination of 2-indexed terms of type 
$C\times C$ contracted by the Minkowskian metric $\eta_{ab}$.
Consider all the possible combinations of the indices and 
take into account the antisymmetry of the $C$-objects to get 
the following list of independent two-indexed terms:
\br 
\label{2-13} 
{}^{(1)}\!A_{ab}&:=&C_{abm}C^m, \qquad {}\\
\label{2-14} 
{}^{(2)}\!A_{ab}&:=&C_{mab}C^m \qquad {\rm\ \ antisymmetric \ object,}\\
\label{2-15} 
{}^{(3)}\!A_{ab}&:=&C_{amn}{C_b}^{mn}\qquad{\rm symmetric \ object,}\\
\label{2-16} 
{}^{(4)}\!A_{ab}&:=&C_{amn}{{C^m}_b}^n, \qquad {}\\
\label{2-17} 
{}^{(5)}\!A_{ab}&:=&C_{man}{{C^n}_b}^m\qquad{\rm symmetric \ object,}\\
\label{2-18} 
{}^{(6)}\!A_{ab}&:=&C_{man}{{C^m}_b}^n\qquad{\rm symmetric \ object,}\\
\label{2-19} 
{}^{(7)}\!A_{ab}&:=&C_aC_b\qquad{\rm\ \ \ \  \ \ \ symmetric \ object.}
\er
In addition to the 2-indexed $A$-objects the general field equation may also include their traces multiplied by $\eta_{ab}$.
These traces of  2-indexed objects are  scalar $SO(1,3)$ invariants:
\br
\label{2-20} 
{}^{(1)}\!A&:=& {}^{(1)}{A_a}^a=-{}^{(7)}{A_a}^a,\\
\label{2-21} 
{}^{(2)}\!A&:=& {}^{(3)}{A_a}^a={}^{(6)}{A_a}^a,\\
\label{2-22} 
{}^{(3)}\!A&:=& {}^{(4)}{A_a}^a={}^{(5)}{A_a}^a.
\er
The trace of the antisymmetric object ${}^{(2)}\!A_{ab}$ is zero.\\
It is easy to see that not all of the objects ${}^{(i)}\!A_{ab},{}^{(i)}\!B_{ab}$ 
are independent. Starting from the relation
$$dd\vt^a=0$$
we obtain:
\br
d({C^a}_{mn}\vt^{mn})&=&{B^a}_{mnp}\vt^{mnp}+{C^a}_{mn}\d\vt^m\vt^n=
({B^a}_{knp}+{C^a}_{mn}{C^m}_{pk})\vt^{pkn}=0.\nonumber
\er
This equation results in a 4-indexed Bianchi identity of the first-order
\begin{equation}\label{2-23}
\boxed{{B^a}_{[knp]}+{C^a}_{m[n}{C^m}_{pk]}=0},
\end{equation}
where $[knp]$ and $[npk]$ are the antisymmetrization of the respective indices.
Taking the unique non-vanishing contraction of the relation (\ref{2-23}) 
we obtain 
$${B^a}_{[kna]}+{C^a}_{m[n}{C^m}_{ak]}=0.$$
Using the antisymmetry of the $B$-objects in the middle indices the 
first term in the l.h.s. of this relation results in  
\br
{B^a}_{[kna]}&=&2\Big({B^a}_{kna}+{B^a}_{nak}+{B^a}_{akn}\Big)=
2\Big(2 \ {}^{(2)}\!B_{[kn]}+{}^{(3)}\!B_{kn}\Big).\nonumber
\er
As for the second part 
\br
{C^a}_{m[n}{C^m}_{ak]}&=&2\Big({C^a}_{mn}{C^m}_{ak}+{C^a}_{ma}{C^m}_{kn}+
{C^a}_{mk}{C^m}_{na}\Big)=2 \ {}^{(2)}\!A_{kn}.\nonumber
\er
Thus (\ref{2-23}) reduces to a 2-indexed Bianchi identity of the first-order 
\begin{equation}\label{2-24}
\boxed
{2\cdot{}^{(2)}\!B_{[kn]}+{}^{(3)}\!B_{kn}+{}^{(2)}\!A_{kn}=0}.
\end{equation}
${}^{(2)}\!B_{[kn]}$ is a linear combination of ${}^{(3)}\!B_{kn}$ and 
${}^{(2)}\!A_{kn}$. Consequently, $\b_6$ in (\ref{2-12}) can be taken to be zero.\\  
The general quadratic part of an equation can be written as
\br\label{2-25}
Q_{ab}&=&\a_1 \ {}^{(1)}\!A_{(ab)}+\a_2 \ {}^{(2)}\!A_{ab}+\a_3 \ {}^{(3)}\!A_{ab}+
\a_4 \ {}^{(4)}\!A_{(ab)}+\a_5 \ {}^{(5)}\!A_{ab}+\nonumber \\
&&\a_6 \ {}^{(6)}\!A_{ab}+\a_7 \ {}^{(7)}\!A_{ab}+\a_8 \ {}^{(1)}\!A_{[ab]}+
\a_9 \ {}^{(4)}\!A_{[ab]}+\nonumber \\
&&
\eta_{ab}\Big(\a_{10} \ {}^{(1)}\!A+\a_{11} \ {}^{(2)}\!A+\a_{12} \ {}^{(3)}\!A\Big),
\er
where $\a_i, \ i=1,...,12$ are free dimensionless  parameters.\\
Let us summarize the construction by
\thm
The most general 2-indexed system of equations satisfying the following conditions 
\begin{itemize}
\item[1.]diffeomorphic covariant  and $SO(1,3)$-invariant, 
\item[2.]linear in the second order derivatives and quadratic in the first order derivatives with coefficients depending on the coframe variables, 
\item[3.]obtained from the quantities ${C^a}_{bc}$ and ${B^a}_{bcd}$ by contractions and transpose
\end{itemize}
is
\begin{equation}\label{2-26}
L_{ab}+Q_{ab}=0,
\end{equation}
where the linear leading part $L_{ab}$ is defined by (\ref{2-12}) 
while the quadratic part $Q_{ab}$ is 
defined by (\ref{2-25}).
\ethm
The field equation (\ref{2-26}) is a system of 16 equations and it can be 
$SO(1,3)$ invariantly decomposed into three independent equations:\\
{\textsl {The trace equation}}
\begin{equation}\label{2-27}
{L_a}^a+{Q_a}^a=0,
\end{equation}
{\textsl {The traceless symmetric equation}}
\begin{equation}\label{2-28}
\Big(L_{(ab)}-\frac 14 L_{m}^m\eta_{ab}\Big)+
\Big(Q_{(ab)}-\frac 14 Q_{m}^m\eta_{ab}\Big)=0,
\end{equation}
{\textsl {And the antisymmetric equation}}
\begin{equation}\label{2-29}
L_{[ab]}+Q_{[ab]}=0.
\end{equation}
Again, the brackets $(ab)$, $[ab]$ mean, respectively,  symmetrization 
and antisymmetrization. \\
The trace equation (\ref{2-27}) can be explicitly written as 
\begin{eqnarray}\label{2-30}
&&(\b_1+\b_2+4\b_4)B+(\a_1-\a_7+4\a_{10}) \ {}^{(1)}\!A+(\a_3+\a_6+4\a_{11}) \ {}^{(2)}\!A+\nonumber\\
&&\qquad(\a_4+\a_5+4\a_{12}) \ {}^{(3)}\!A=0.
\end{eqnarray}
As for the traceless symmetric equation 
\begin{eqnarray}\label{2-31}
\b_1 \ \overline{{}^{(1)}\!B_{(ab)}}+\b_2 \ \overline{{}^{(2)}\!B_{(ab)}}&+&
\a_1 \ \overline{{}^{(1)}\!A_{(ab)}}+\a_3 \ \overline{{}^{(3)}\!A_{ab}}+
\a_4 \ \overline{{}^{(4)}\!A_{(ab)}}\nonumber\\
&+&
\a_5 \ \overline{{}^{(5)}\!A_{ab}}+
\a_6 \ \overline{{}^{(6)}\!A_{ab}}+\a_7 \ \overline{{}^{(7)}\!A_{ab}}=0,
\end{eqnarray}
where the bar means
\begin{equation}\label{2-32}
\overline {M_{ab}}=M_{ab}-\frac 14\eta_{ab} {M_{m}}^m.
\end{equation}
The antisymmetric equation is
\begin{eqnarray}\label{2-33}
&&
\b_3 \ {}^{(3)}\!B_{ab}+\b_5 \ {}^{(1)}\!B_{[ab]}+\b_6 \ {}^{(2)}\!B_{[ab]}+
\a_2 \ {}^{(2)}\!A_{ab}+\a_8 \ {}^{(1)}\!A_{[ab]}+\a_9 \ {}^{(4)}\!A_{[ab]}=0.
\end{eqnarray}
This way a general  family of field equations for the coframe field $\vt^a$ was constructed. Every equation in the family is invariant under diffeomorphic transformations of the coordinate system.  It is invariant  under global $SO(1,3)$-transformations of the coframe as well. 
In the following sections we impose additional two conditions:
\begin{itemize}
\item An   action condition. That means that the equation is  derivable from a suitable action.   
\item  Long-distance approximation conditions. That means that the equation has a solution  confirming with the observed data.
\end{itemize}

\sect{Quadratic Lagrangians}
One of the basic tools to derive field equations is the variational principle. 
By that a suitable Lagrangian is chosen and it's variation is equated to zero. 
In the  teleparallel approach the coframe field $\vt^a$ is the basic 
 field variable, while the Lagrangian $\L$ is a differential 4-form.
A general Lagrangian density for the coframe field $\vt^a$ (quadratic in 
the first order derivatives and linear in the second order derivatives) can be 
expressed as a linear combination  of scalar $A$- and $B$-objects.
\begin{equation}
\L=\frac 1{\ell^2}\Big(\mu_0 \  B+\mu_1 \ {}^{(1)}\!A+\mu_2 \ {}^{(2)}\!A+
\mu_3 \ {}^{(3)}\!A\Big)*1,
\label{3-1}
\end{equation}
where ${\ell}$ is a  length-dimensional constant, $B$ is the second order 
scalar object, defined by (\ref{2-9}) and ${}^{(i)}\!A$ with $i=1,2,3$ are quadratic scalar objects defined by (\ref{2-20}-\ref{2-22}).\\
The terms in the expression (\ref{3-1}) are completely independent. They are 
 diffeomorphic covariant and rigid $SO(1,3)$ invariant.
Let us show that the linear combination (\ref{3-1}) is equivalent  
to the gauge invariant translation Lagrangian of Rumpf \cite{Rumpf}
(up to his $\Lambda$-term).
\begin{equation}\label{3-2}
V=\frac 1{2\ell^2}\sum^3_{I=1}\rho_I{}^{(I)}V,
\end{equation}
where
\br\label{3-3}
{}^{(1)}V&=&d\vt^a\wedge*d\vt_a,\\
\label{3-4}
{}^{(2)}V&=&\Big(d\vt_a \wedge \vt^a \Big) \wedge*( d\vt_b \wedge \vt^b),\\
\label{3-5}
{}^{(3)}V &=&(d\vt_a \wedge\vt^b ) \wedge * \Big(d\vt_b \wedge \vt^a \Big).
\er
Indeed, the first term (\ref{3-3}) can be rewritten as  
\br \label{3-6}
{}^{(1)}V&=&{C^a}_{mn}\vt^{mn}\wedge*{C_a}^{pq}\vt_{pq}
=2C_{amn}C^{amn}*1=2 \  {}^{(2)}\!A*1.
\er
As for the second term (\ref{3-4}) 
\br
\label{3-7}
{}^{(2)}V&=&C_{amn}\vt^{amn}\wedge C_{bpq}*\vt^{bpq}=
-2C_{amn}(C^{amn}+C^{mna}+C^{nam})*1\nonumber\\
&=&2(2 \  {}^{(3)}\!A-{}^{(2)}\!A)*1.
\er
The third term (\ref{3-5}) takes the form 
\br
\label{3-8}
{}^{(3)}V&=&C_{amn}\vt^{mnb}\wedge C_{bpq}*\vt^{pqa}=-4C_mC^m*1=
2(2  \  {}^{(1)}-A{}^{(2)}\!A)*1.
\er
Thus the coefficients of (\ref{3-1}) are the linear combinations of the 
coefficients of (\ref{3-2}).\\
As for the second derivative term $B$, 
\brn
B*1&=&{{B^a}_{ab}}^b*1=\eta^{bc}{B^a}_{abc}*1=
\eta^{bc}(e_c\hook d{C^a}_{ab})*1\\
&=&-\vt^b\wedge *d({C^a}_{ab})=-d({C^a}_{ab})\wedge *\vt^b\\
&=&-d({C^a}_{ab} *\vt^b)+{C^a}_{ab}d*\vt^b=d(C_b *\vt^b)-C_b d*\vt^b.
\ern
Using the relation 
\begin{equation}
\label{3-9}
d*\vt^b=-C^b*1
\end{equation}
we obtain
\begin{equation}\label{3-10}
B*1=d(C_a *\vt^a)+C_{a}C^{a} *1=d(C_a *\vt^a)-{}^{(1)}\!A*1.
\end{equation}
It is well known that  total derivatives do not contribute to the field 
equation. 
Thus we can neglect the $B$-term in the Lagrangian (\ref{3-1}). \\
The comparison of (\ref{3-1}) and (\ref{3-2}) yields
\begin{equation}\label{3-11}
\boxed{
\mu_1=2\rho_3, \qquad \mu_2=\rho_1-\rho_2-\rho_3, \qquad \mu_3=2\rho_2}.
\end{equation}
Therefore, the Rumpf Lagrangian (\ref{3-2}) is the most general 
quadratic Lagrangian. \\
In \cite{Hehl}, the Einsteinian theory of gravity is 
reinstated from the 
Lagrangian (\ref{3-2}) by letting 
\begin{equation}\label{3-12}
\rho_1=0, \qquad  \rho_2=-\frac 12, \qquad  \rho_3=1.
\end{equation}
Thus, correspondingly 
\begin{equation}\label{3-13}
\mu_1=2, \qquad  \mu_2=-\frac 12, \qquad  \mu_3=-1.
\end{equation}
This way we have shown that the general translation invariant Lagrangian can 
be expressed by the scalar $A$-objects.
\sect{The action generated field equation}
It is natural to expect that the field equation, derived from the Lagrangian above can be expressed in terms of the $A$ and $B$-objects. 
The free variations of the Rumpf Lagrangian (\ref{3-2}) yield the field equation due to Kopczy\'nski
\cite{Kop}. Let us express it  in the following form (Cf. \cite{Hehl}):
\br\label{4-1}
  - 2\ell^2\Sigma_a &=& 2\rho_1 d*d\vt_a
 - 2\rho_2 \vt_a \wedge d *(d\vt^b\wedge \vt_b)-
  2\rho_3\vt_b \wedge d*( \vt_a \wedge d \vt^b ) \nonumber \\
 && + \rho_1 \Big[ e_a \hook( d\vt^b \wedge * d\vt_b) 
 -2 ( e_a \hook d\vt^b ) \wedge *d\vt_b\Big] \nonumber \\
 && + \rho_2\Big[2 d\vt_a \wedge * ( d\vt^b \wedge \vt_b) 
 + e_a \hook ( d\vt^c \wedge\vt_c \wedge * ( d\vt^b \wedge\vt_b) )\nonumber\\
&&\qquad -2( e_a \hook d\vt^b) \wedge \vt_b \wedge *
  ( d\vt^c \wedge \vt_c)\Big] \nonumber \\
 && +\rho_3\Big[2 d\vt_b \wedge * ( \vt_a\wedge d \vt^b)
+ e_a \hook( \vt_c \wedge d\vt^b \wedge *( d\vt^c \wedge \vt_b )) \nonumber \\
&&\qquad
- 2( e_a \hook d\vt^b) \wedge \vt_c \wedge * ( d\vt^c \wedge\vt_b )\Big] ,
\er
where $\Sigma_a$ depends on matter fields.\\ 
By  Appendix A this equation can be rewritten  as
\br\label{4-2}
- 2\ell^2\Sigma_a &=&\rho_1\Big(-2 \ {}^{(1)}\!B_{ab}-2 \ {}^{(1)}\!A_{ab}-{}^{(3)}\!A_{ab}
-\frac 12  \ {}^{(2)}\!A\eta_{ab}+4 \ {}^{(6)}\!A_{ab}\Big)*\vt^b\nonumber\\
&&\rho_2\Big(4 \ {}^{(1)}\!B_{[ab]}+2 \ {}^{(3)}\!B_{ab}
+4 \ {}^{(1)}\!A_{[ab]}
+2 \ {}^{(2)}\!A_{[ab]}+
3 \ {}^{(3)}\!A_{ab}\nonumber\\&&\qquad
+\frac 12{}^{(2)}\!A\eta_{ab}
-{}^{(3)}\!A\eta_{ab}
+2 \ {}^{(5)}\!A_{ab}
-2 \ {}^{(6)}\!A_{ab}
\Big)*\vt^b+\nonumber\\
&&\rho_3\Big(
2 \ {}^{(1)}\!B_{ab}+2 \ {}^{(2)}\!B_{ba}-2B\eta_{ab}+
2 \ {}^{(1)}\!A_{ab}
-3 \ {}^{(1)}\!A\eta_{ab}\nonumber\\&&\qquad
+\frac12 \ {}^{(2)}\!A\eta_{ab}
+{}^{(3)}\!A_{ab}-
2 \ {}^{(6)}\!A_{ab}
\Big)*\vt^b.
\er
Observe, that, of all the objects defined in (\ref{2-26}) only the objects ${}^{(2)}\!A_{(ab)}$ and ${}^{(4)}\!A_{ab}$ 
are missing in (\ref{4-2}). \\
The calculations above can be summarized by
\thm
The field equation generated by the variation of Rumpf Lagrangian are 
expressed by combination of the structural $A$ and $B$-objects as  
in (\ref{4-2}).
\ethm
Consider the special case, when the antisymmetric part of  equation 
(\ref{4-2}) is identically zero.
Extracting the antisymmetric part of equation (\ref{4-2}) we obtain
\br\label{4-3}
&&
\ell^2*(\vt_a\wedge\Sigma_b-\vt_b\wedge\Sigma_a) =
\rho_1\Big(-2 \ {}^{(1)}\!B_{[ab]}-2 \ {}^{(1)}\!A_{[ab]}\Big)+\nonumber\\
&&\qquad\rho_2\Big(4 \ {}^{(1)}\!B_{[ab]}+2 \ {}^{(3)}\!B_{[ab]}
+4 \ {}^{(1)}\!A_{[ab]}
+2 \ {}^{(2)}\!A_{[ab]}
\Big)+\nonumber\\
&&\qquad\rho_3\Big(
2 \ {}^{(1)}\!B_{[ab]}+2 \ {}^{(2)}\!B_{[ba]}+
2 \ {}^{(1)}\!A_{[ab]}
\Big).
\er
Impose the symmetry condition on the matter current $\Sigma_a$
\begin{equation} \label{4-4}
\vt_a\wedge\Sigma_b=\vt_b\wedge\Sigma_a.
\end{equation} 
Substitute the Bianchi identity (\ref{2-24}), to get
\br\label{4-5}
&&(-2\rho_1+4\rho_2+2\rho_3) \ {}^{(1)}\!B_{[ab]}+
(2\rho_2+\rho_3) \ {}^{(3)}\!B_{[ab]}+\nonumber\\
&&\qquad(-2\rho_1+4\rho_2+2\rho_3) \ {}^{(1)}\!A_{[ab]}+
(2\rho_2+\rho_3) \ {}^{(2)}\!A_{[ab]}=0.
\er
The l.h.s. of this equation is identically zero if and only if
\begin{equation}\label{4-6}
\rho_1=0,  \qquad 2\rho_2+\rho_3=0.
\end{equation}
By homogeneity  we obtain that the system (\ref{4-6}) is equivalent to the system (\ref{3-12}). 
This is the case for the teleparallel equivalent of the 
Einsteinian gravity. In this case the metric is an independent field variable 
and the field equation is restricted to  a system of 10 independent equations. 
Thus we have shown that  Einstein equation is the unique symmetric field 
equation that can be derived from a quadratic Lagrangian. 
\sect{Diagonal static ansatz}
Another subclass of the general field equation (\ref{2-26}) can be constructed by 
with the requirement to have a solution which is confirmed by the observational  
data. We restrict ourselves to the three classical gravity tests, 
namely the Mercury perihelion shift, the light ray shift and the red shift. 
The experimental results can be described by a metric element  
$$
ds^2=-Fdt^2+G(dx^2+dy^2+dz^2),
$$
with
\brn
F&=&1-\frac{2m}{r}+\frac{4m^2}{r^2}+\cdots,\\
G&=&1+\frac{2m}{r}+\cdots.
\ern
The contribution of the third order term in the temporal component and the second order term in the spatial component can not be experimentally detected.\\
In order to obtain a metric of such type we begin with a diagonal static ansatz
\begin{equation}\label{5-1}
\vt^0=e^fdx^0,\qquad \vt^m=e^gdx^m,\qquad m,n=1,2,3,
\end{equation}
where $f$ and $g$ are two arbitrary functions of the spatial coordinates
$x,y,z$.
Substitute of (\ref{5-1}) into equation (\ref{2-26}) to get 
\thm
All the possible solutions of the equation 
$$L_{ab}+Q_{ab}=0$$
of the form (\ref{5-1}) are determined by the solutions of 
\begin{equation}\label{5-2}\boxed{
\mu_1\triangle f+\mu_2\triangle g=\mu_3(\nabla f \ \nabla f)+
\mu_4(\nabla g \ \nabla g)+
\mu_5(\nabla f \ \nabla g)}.
\end{equation}
\begin{equation}\label{5-3}\boxed{
\begin{array}{ll}
&\eta_{mn}(\nu_1\triangle f+\nu_2\triangle g)+\nu_3 f_{mn}+\nu_4g_{mn}=
\\&
\qquad \qquad\eta_{mn}(\nu_5(\nabla f \ \nabla f)+\nu_6(\nabla g \ \nabla g)+\nu_7(\nabla f \ \nabla g))+\\
&\qquad \qquad  \nu_8 f_mf_n+\nu_9g_mg_n+\nu_{10}f_mg_n+\nu_{11}f_ng_m,
\end{array}}
\end{equation}
where $\mu_1,...,\mu_5$ and $\nu_1,...,\nu_{11}$ 
determined by
\br\label{5-4}
\mu_1&=&\b_1+\b_4, \nonumber\\
\mu_2&=&2\b_4, \nonumber\\
\mu_3&=&\a_1-2\a_3-\a_4-\a_6+\a_{10}-2\a_{11}-\a_{12}, \nonumber\\
\mu_4&=&2\b_4+4\a_{10}-4\a_{11}-2\a_{12},\nonumber\\
\mu_5&=&\b_1+\b_4+2\a_1+4\a_{10}, \nonumber\\
\nu_1&=&\b_4,  \nonumber\\
\nu_2&=&\b_1+2\b_4,  \nonumber\\
\nu_3&=&0, \\ 
\nu_4&=&\b_1-2\b_2,\nonumber\\
\nu_5&=&\a_{10}-2\a_{11}-\a_{12},  \nonumber\\
\nu_6&=&\b_1+2\b_4+2\a_1-2\a_3-\a_4-\a_6+4\a_{10}-4\a_{11}-2\a_{12}, \nonumber\\
\nu_7&=&\b_4+\a_1+4\a_{10},  \nonumber\\
\nu_8&=&\a_5+\a_6+\a_7,\nonumber\\
\nu_9&=&-\b_1+\b_2+2\a_1-2\a_3-\a_4+2\a_5+\a_6+4\a_7,  \nonumber\\
\nu_{10}&=&-\b_2-\a_2+2\a_7, \nonumber\\ 
\nu_{11}&=&\a_2+\a_7.\nonumber
\er
\ethm
The detailed computations are carried in Appendix B. 
Only 12 coefficients in (\ref{5-4}) are independent. Indeed, in addition to the 
relations $\nu_3=0$ and $2\nu_1=\mu_2$ the $\mu_i$ and $\nu_i$ have to satisfy 
\begin{equation}\label{5-5}
-2\mu_1+4\mu_3+2\mu_5 +4\nu_2+\nu_5=0
\end{equation}
and
\begin{equation}\label{5-6}
3\mu_1-\mu_2-2\mu_4-\mu_5+2\nu_2+4\nu_5=0.
\end{equation}
Observe that (\ref{5-3}) is a system of 4 equations. Thus, the system (\ref{5-2}) and (\ref{5-3}) is over-determined. This means that the coefficients 
$\mu_k$ and $\nu_k$ have to be chosen so that two independent equation for 
$f$ and $g$ are left. This can be done in a variety of ways. \\
Let us turn to a special case of the spherical symmetry.
\thm
If $f$ and $g$ are functions of the radial coordinate $r=(x^2+y^2+z^2)^{1/2}$ 
then  (\ref{5-2}) and (\ref{5-3}) read
\begin{equation}\label{5-7}
\left\{\begin{array}{ll}
& \mu_1f''+\mu_2g''+2\frac 1r(\mu_1f'+\mu_2g')=\mu_3{f'}^2+\mu_4{g'}^2+\mu_5f'g' \\
&\\
& \nu_1f''+\nu_2g''+\frac 1r[2\nu_1f'+(2\nu_2+\nu_4)g']=
\nu_5{f'}^2+\nu_6{g'}^2+\nu_7f'g' \\
&\\
& \nu_4g''+\frac {1}{r}\nu_4g'=\nu_8 {f'}^2+
\nu_9{g'}^2+(\nu_{10}+\nu_{11})f'g'
\end{array}\right.
\end{equation}
\ethm
This is obtained by a direct substitution of spherical-symmetric ansatz in (\ref{5-2}) and (\ref{5-3}). Observe that in this case 
the terms $f_{mn}$, $g_{mn}$, $f_mf_n$, $g_mg_n$ and $f_mg_n$ all contain $x_mx_n$ as a 
factor. Thus the equation (\ref{5-3}) is decomposed into two distinct equations. Note that the system (\ref{5-7}) is still  over-determined - three ODE for two independent variables $f$ and $g$. An obvious way to reduce (\ref{5-7}) to two equations is to take 
$\nu_4=\nu_8=\nu_9=\nu_{10}+\nu_{11}=0$.
\sect{Approximate solutions}
In order to confirm the field equation  (\ref{2-26}) with the observed data 
we construct an approximate solution of the restricted system (\ref{5-7}).
 Correspondently we consider the 
long-distance approximation of the functions $f$ and $g$. This means that the weak field on a distance far greater than the mass of the body (in the natural system of units) is studied. Take the Taylor expansion of the functions $f$ and $g$:
\br\label{6-1}
f&=&1+\frac {a_1}{r}+\frac {a_2}{r^2}+ \cdots,\\
\label{6-2}
g&=&1+\frac {b_1}{r}+\frac {b_2}{r^2}+ \cdots.
\er
So, the first equation of the system (\ref{5-7}) takes the form
\brn
&&\mu_1(6\frac {a_2}{r^4}+ \cdots)
+\mu_2(6\frac {b_2}{r^4}+ \cdots)
+2\frac 1r\Big(\mu_1(-2\frac {a_2}{r^3}+ \cdots)
+\mu_2(-2\frac {b_2}{r^3}+ \cdots)\Big)=\\
&&\mu_3{(\frac {a_1}{r^2}+2\frac {a_2}{r^3}+ \cdots)}^2
+\mu_4{(\frac {b_1}{r^2}+2\frac {b_2}{r^3}+ \cdots)}^2
+\mu_5(\frac {a_1}{r^2}+2\frac {a_2}{r^3}+ \cdots)(\frac {b_1}{r^2}+2\frac {b_2}{r^3}+ \cdots).
\ern
Thus, up to $O(\frac 1{r^4})$
\begin{equation}\label{6-3}
2\mu_1a_2+2\mu_2b_2=
\mu_3a_1^2+\mu_4b_1^2+\mu_5a_1b_1.
\end{equation}
As for the second equation of the system (\ref{5-7}) 
\brn
&&\nu_1(6\frac {a_2}{r^4}+ \cdots)
+\nu_2(6\frac {b_2}{r^4}+ \cdots)+
\frac 1r[2\nu_1(-2\frac {a_2}{r^3}+ \cdots)
+(2\nu_2+\nu_4)(-2\frac {b_2}{r^3}+ \cdots)]\\
&=&
\nu_5{(\frac {a_1}{r^2}2\frac {a_2}{r^3}+ \cdots)}^2
+\nu_6{(\frac {b_1}{r^2}+2\frac {b_2}{r^3}+ \cdots)}^2+
\nu_7(\frac {a_1}{r^2}+2\frac {a_2}{r^3}+ \cdots)(\frac {b_1}{r^2}+2\frac {b_2}{r^3}+ \cdots).
\ern
Thus 
\begin{equation}\label{6-4}
\nu_4b_1=0,
\end{equation}
\begin{equation}\label{6-5}
2\nu_1a_2+2\nu_2b_2-2\nu_4b_2=\nu_5a_1^2+\nu_6b_1^2+\nu_7a_1b_1,
\end{equation}
The third equation of the system (\ref{5-7}) is
\brn
 &&\nu_4(2\frac {b_1}{r^3}+6\frac {b_2}{r^4}+ \cdots)
+\frac {1}{r}\nu_4(-\frac {b_1}{r^2}-2\frac {b_2}{r^3}+ \cdots)=
\nu_8 {(-\frac {a_1}{r^2}-2\frac {a_2}{r^3}+ \cdots)}^2+\\&&
\nu_9{(-\frac {b_1}{r^2}-2\frac {b_2}{r^3}+ \cdots)}^2+
(\nu_{10}+\nu_{11})(-\frac {a_1}{r^2}-2\frac {a_2}{r^3}+ \cdots)
(-\frac {b_1}{r^2}-2\frac {b_2}{r^3}+ \cdots).
\ern
Consequently
\begin{equation}\label{6-6}
\nu_4b_1=0
\end{equation}
\begin{equation}\label{6-7}
4\nu_4b_2=\nu_8a_1^2+\nu_9b_1^2+
(\nu_{10}+\nu_{11})a_1b_1.
\end{equation}
The classical gravity tests (Mercury perihelion shift, light ray shift 
and the red shift) can be described by the following choice of the coefficients
\begin{equation}\label{6-8}
a_1=-m,\qquad b_1=m,\qquad a_2=m^2,\qquad b_2=km^2,
\end{equation}
where $m$ is the mass of the Sun in dimensionless units. As for $k$, it is an arbitrary 
dimensionless constant. This really means that $\b_2$ is free.
For such choice of it follows that 
\begin{equation}\label{6-9}
\left\{\begin{array}{ll}
&\nu_4=0,\\
&2\mu_1+2k\mu_2=\mu_3+\mu_4-\mu_5,\\
&2\nu_1+2\nu_2k=\nu_5+\nu_6-\nu_7,\\
&\nu_8+\nu_9=\nu_{10}+\nu_{11}
\end{array}\right.
\end{equation}
By (\ref{5-4}) rewrite these relations by the coefficients $\a_i,\b_k$.
\begin{equation}\label{6-10}
\left\{\begin{array}{ll}
& \b_1=2\b_2, \\
&6\b_2+(1+4k)\b_4=
-\a_1-2\a_3-\a_4-\a_6+\a_{10}-6\a_{11}-3\a_{12},\\
&2\b_2(2k-1)+(4k+1)\b_4=\a_1-2\a_3-\a_4-\a_6+\a_{10}-6\a_{11}-3\a_{12},\\
&2\a_1-2\a_3-\a_4+3\a_5+3\a_6+2\a_7=0.
\end{array}\right.
\end{equation}
Or
\begin{equation}\label{6-11}
\left\{\begin{array}{ll}
& \b_1=2\b_2, \\
&2(k-2)\b_2=\a_1,\\
&6\b_2+(1+4k)\b_4=-\a_1-2\a_3-\a_4-\a_6+\a_{10}-6\a_{11}-3\a_{12},\\
&2\a_1-2\a_3-\a_4+3\a_5+3\a_6+2\a_7=0
\end{array}\right.
\end{equation}
In order to eliminate the arbitrary constant $k$ from the system (\ref{6-11}) 
we first consider the special case
\begin{equation}\label{6-12}
\b_2=0.
\end{equation}
The system (\ref{6-11}) takes now the form
\br\label{6-13}
&&\b_1=\b_2=\a_1=0,\\
&&(1+4k)\b_4=-2\a_3-\a_4-\a_6+\a_{10}-6\a_{11}-3\a_{12},\\
\label{6-14}
&&-2\a_3-\a_4+3\a_5+3\a_6+2\a_7=0.
\label{6-15}
\er
The equation (\ref{6-14}) gives no information, since it contains the arbitrary constant $k$. The remaining equations (\ref{6-13}) and (\ref{6-15}) do not constitute  a viable physical system. 
Indeed, in this case, the traceless symmetric equation (\ref{2-31}) in this case has no 
a leading (second derivatives) part. Thus one can not get an approximation by a wave equation for 
small fluctuations of the coframe and consequently of the metric tensor.\\
 In the case $\b_2\ne 0$ we use the homogeneity of the system (\ref{6-11}) 
to take $\b_2=1$. 
Thus the system (\ref{6-11}) is rewritten  
\begin{equation}\label{6-18}
\left\{\begin{array}{ll}
& \b_1=2, \\
& \b_2=1, \\
&(9+2\a_1)\b_4=-6-\a_1-2\a_3-\a_4-\a_6+\a_{10}-6\a_{11}-3\a_{12},\\
&2\a_1-2\a_3-\a_4+3\a_5+3\a_6+2\a_7=0
\end{array}\right.
\end{equation}
\thm
Any operator that satisfies (\ref{6-18}) have solutions that can not be experimentally distinguished, 
by the three classical tests, from Schwarzschild solution.
\ethm
This way we obtain a wide class of field equations which are confirmed 
by three classical tests.
\section*{Acknowledgment}
Authors are grateful to Prof. F.\,W.~Hehl for useful discussions and valuable comments.
\section*{Appendices}
\appendix
\sect{Field equation from action}
In order to rewrite the field equation (\ref{4-1}) in terms of 
the $A,B,C$-variables we use the following formulas
\br
\label{formula1}
d^+(\vt^a)&=&C^a,\\
\label{formula2}
d^+(\vt^{ab})&=&C^a\vt^b-C^b\vt^a-{C_m}^{ab}\vt^m,\\
\label{formula3}
d^+(\vt^{abc})&=&C^a\vt^{bc}-C^b\vt^{ac}+C^c\vt^{ab}+
{C_m}^{bc}\vt^{am}-{C_m}^{ac}\vt^{bm}+{C_m}^{ab}\vt^{cm},\\
\label{formula4}
d^+(\vt^{abcd})&=&0.
\er
For the first second derivative term in equation (\ref{4-1}) we obtain
\br
\boxed{2\rho_1d*d\vt_a}&=&\rho_1d\Big(C_{abc}*\vt^{bc}\Big)=
\rho_1{B_{abc}}^m\vt_m\wedge *\vt^{bc}+\rho_1C_{abc}d*\vt^{bc}\nonumber \\
&=&\rho_1{B_{abc}}^m(\d_m^b*\vt^c-\d_m^c*\vt^b)+\rho_1C_{abc}d*\vt^{bc}\nonumber \\
&=&2\rho_1{B_{amc}}^m*\vt^c+\rho_1C_{abc}(C^b*\vt^c-C^c*\vt^b-{C_m}^{bc}*\vt^m)\nonumber \\
&=&\rho_1(2{B_{amc}}^m+C_{abc}C^b-C_{acb}C^b-C_{abm}{C_c}^{bm})*\vt^c\nonumber \\
&=&\boxed{-\rho_1\Big(2 \ {}^{(1)}\!B_{ab}+2 \ {}^{(1)}\!A_{ab}+{}^{(3)}\!A_{ab}\Big)*\vt^b}
\er
The second term in (\ref{4-1}) takes the form
\vspace{0in}
\br
&&\boxed{-2\rho_2\vt_a \wedge d *(d\vt_b\wedge \vt^b)}=
-\rho_2\vt_a \wedge d(C_{bmn}*\vt^{mnb})\nonumber \\&&
=-\rho_2\Big({B_{bmn}}^k\vt_{ak}\wedge*\vt^{mnb}
+C_{bmn}\vt_a \wedge d*\vt^{mnb}\Big)\nonumber \\&&=
\rho_2{B_{bmn}}^k\vt_a\wedge *(\d^m_k\vt^{nb}-\d^n_k\vt^{mb}+\d^b_k\vt^{mn})+\nonumber \\&&\quad
\rho_2C_{bmn}\vt_a \wedge*\Big[C^m\vt^{nb}-C^n\vt^{mb}+C^b\vt^{mn}
+\Big({C_k}^{nb}\vt^{mk}-{C_k}^{mb}\vt^{nk}+{C_k}^{mn}\vt^{bk}\Big)\Big]
\nonumber \\&&=
\rho_2{B_{bmn}}^k*\Big(\d^m_k(\d^n_a\vt^{b}-\d^b_a\vt^{n})-
\d^n_k(\d^m_a\vt^{b}-\d^b_a\vt^{m})
+\d^b_k(\d^m_a\vt^{n}-\d^n_a\vt^{m})\Big)
\nonumber \\&&\quad 
\rho_2C_{bmn}*\Big[C^m(\d^n_a\vt^{b}-\d^b_a\vt^{n})+
C^n(\d^m_a\vt^{b}-\d^b_a\vt^{m})+C^b(\d^m_a\vt^{n}-\d^n_a\vt^{m})
\nonumber \\&&
\quad -\rho_2\Big(
{C_k}^{nb}(\d^m_a\vt^{k}-\d^k_a\vt^{m})-
{C_k}^{mb}(\d^n_a\vt^{k}-\d^k_a\vt^{n})
+{C_k}^{mn}(\d^b_a\vt^{k}-\d^k_a\vt^{b})
\Big)\Big]\nonumber \\&&=
2\rho_2({B_{nka}}^k+{B_{ank}}^k+{B_{kan}}^k)*\vt^{n}+
2\rho_2C^m(C_{bma}+C_{abm}+C_{mab})*\vt^{b}\nonumber \\&&\quad
+2\rho_2\Big(C_{man}{C_k}^{nm}+C_{amn}{C_k}^{mn}+C_{amn}{C^{mn}}_k\Big)*\vt^k
\nonumber \\&&=
\boxed{
\rho_2\Big(4 \ {}^{(1)}\!B_{[ab]}+2 \ {}^{(3)}\!B_{ab}+4 \ {}^{(1)}\!A_{[ab]}
+2 \ {}^{(2)}\!A_{[ab]}+
2 \ {}^{(3)}\!A_{ab}+4 \ {}^{(4)}\!A_{[ab]}\Big)*\vt^b}\nonumber \\&&
\er
Note, that the first five terms in the brackets are antisymmetric matrices 
while the last one, namely ${}^{(3)}\!A_{ab}$, is symmetric.\\
The third term in equation (\ref{4-1}) is
\br
&&\!\!\!\!\!\!\!\!\!\boxed{-2\rho_3\vt_b \wedge d*( \vt_a \wedge d \vt^b )}=-\rho_3
\vt_b \wedge d\Big(C^{bmn}*\vt_{amn}\Big)=\nonumber\\
&&\!\!\!\!\!\!\!\!\!
-\rho_3B^{bmnk}\vt_{bk}\wedge*\vt_{amn}-\rho_3C^{bmn}\vt_b \wedge d*\vt_{amn}\nonumber\\
&&\!\!\!\!\!\!\!\!\!=
\rho_3{{B_b}^{mn}}_k*[e^b\hook(e^k\hook\vt_{amn})]+\rho_3C^{bmn}*(e_b \hook d^+\vt_{amn})
\nonumber\\
&&\!\!\!\!\!\!\!\!\!
=\rho_3{{B_b}^{mn}}_k*\Big(e^b\hook(\d^k_a\vt_{mn}-\d^k_m\vt_{an}+\d^k_n\vt_{am})\Big)
\nonumber\\
&&\!\!\!\!\!\!\!\!\!\quad+\rho_3{C_b}^{mn}*\Big[e^b \hook\Big( C_a\vt_{mn}-
C_m\vt_{an}+C_n\vt_{am}+
{C^k}_{mn}\vt_{ak}-{C^k}_{an}\vt_{mk}+{C^k}_{am}\vt_{nk}\Big)\Big]
\nonumber\\
&&\!\!\!\!\!\!\!\!\!=\rho_3{{B_b}^{mn}}_k*\Big(\d^k_a(\d^b_m\vt_n-\d^b_n\vt_m)-
\d^k_m(\d^b_a\vt_n-\d^b_n\vt_a)+\d^k_n(\d^b_a\vt_m-\d^b_m\vt_a)\Big)\nonumber\\
&&\!\!\!\!\!\!\!\!\!\quad+\rho_3{C_b}^{mn}*\Big( C_a(\d^b_m\vt_n-\d_n^b\vt_m)
-C_m(\d_a^b\vt_n-\d_n^b\vt_a)+C_n(\d_a^b\vt_m-\d_m^b\vt_a)\Big)\nonumber\\
&&\!\!\!\!\!\!\!\!\!\quad+\rho_3{C_b}^{mn}*\Big({C^k}_{mn}(\d^b_a\vt_k-\d_k^b\vt_a)-
{C^k}_{an}(\d_m^b\vt_k-\d_k^b\vt_m)+{C^k}_{am}(\d_n^b\vt_k-\d_k^b\vt_n)\Big)\nonumber\\
&&\!\!\!\!\!\!\!\!\!=2\rho_3*\Big({{B_b}^{bn}}_a\vt_n+{{B_a}^{mn}}_n\vt_m+{{B_b}^{mb}}_m\vt_a\Big)
2\rho_3*\Big({C_b}^{bn} C_a\vt_n+{C_b}^{mb}C_m\vt_a+{C_a}^{mn}C_n\vt_m\Big)
\nonumber\\
&&\!\!\!\!\!\!\!\!\!\quad+\rho_3*\Big({C_a}^{mn}C_{bmn}-{C_k}^{mn}{C^k}_{mn}
+2C^mC_{bam}+2{C_{kb}}^n{C^k}_{an}\Big)\vt^b
\nonumber\\
&&\!\!\!\!\!\!\!\!\!=\boxed{\rho_3\Big(2 \ {}^{(1)}\!B_{ab}+2 \ {}^{(2)}\!B_{ba}-
2B\eta_{ab}+
4 \ {}^{(1)}\!A_{(ab)}-2 \ {}^{(1)}\!A\eta_{ab}}\nonumber\\
&&\!\!\!\!\!\!\!\!\!\quad\boxed{\quad-
{}^{(2)}\!A\eta_{ab}+{}^{(3)}\!A_{ab}+
2 \ {}^{(6)}\!A_{ab}-2 \ {}^{(7)}\!A_{ab}\Big)*\vt^b}
\er
The first quadratic term in equation (\ref{4-1}) is
\br
\boxed{\rho_1e_a \hook( d\vt^b \wedge * d\vt_b)}&=&
\frac 14 \rho_1{C^b}_{mn}{C_b}^{pq}e_a \hook(\vt^{mn}\wedge *\vt_{pq})
=\frac 14 \rho_1{C^b}_{mn}{C_b}^{pq}(\d^n_p\d^m_q-\d^n_q\d^m_p)*\vt_a\nonumber\\
&=&-\frac 12 \rho_1C_{bmn}C^{bmn}*\vt_a=\boxed{-\frac 12 \rho_1 \ {}^{(2)}\!A\eta_{ab}*\vt^b}
\er
The second quadratic term in equation (\ref{4-1}) is
\br
&&\boxed{-2\rho_1(e_a\hook d\vt^b ) \wedge *d\vt_b}=
-\frac 12 \rho_1{C^b}_{mn}{C_b}^{pq}(e_a\hook\vt^{mn})\wedge *\vt_{pq}
\\&&\qquad 
=-\frac 12 \rho_1{C^b}_{mn}{C_b}^{pq}(\d_a^m\vt^n-\d_a^n\vt^m)\wedge *\vt_{pq}\nonumber\\
&&\qquad =
-\frac 12 \rho_1{C^b}_{mn}{C_b}^{pq}(\d_a^m\d^n_p*\vt_q-\d_a^m\d^n_q*\vt_p
-\d_a^n\d^m_p*\vt_q+\d_a^n\d^m_q*\vt_p)\nonumber\\
&&\qquad = 4\rho_1C_{bap}C^{bqp}*\vt_q=\boxed{4\rho_1 \ {}^{(6)}\!A_{ab}*\vt^b}
\er
The third quadratic term in equation (\ref{4-1}) is
\br
&&\boxed{2\rho_2 d\vt_a \wedge * ( d\vt^b \wedge \vt_b)}=
\frac 12\rho_2C^{bmn}C_{apq}\vt^{pq}\wedge * \vt_{bmn}=
-\frac 12\rho_2C^{bmn}C_{apq}\vt^p\wedge*^2(\vt^q\wedge * \vt_{bmn})\nonumber\\
&&\qquad = -\frac 12\rho_2C^{bmn}C_{apq}\vt^p\wedge(\d^q_b*\vt_{mn}-\d^q_m*\vt_{bn}+
\d^q_n*\vt_{bm})\nonumber\\
&&\qquad = -\frac 12\rho_2C^{bmn}C_{apq}*(\d^q_b\d^p_m\vt_{n}-\d^q_b\d^p_n\vt_{m}
-\d^q_m\d^p_b\vt_{n}+\d^q_m\d^p_n\vt_{b}+\d^q_n\d^p_b\vt_{m}
-\d^q_n\d^p_m\vt_{b})\nonumber\\
&&\qquad =-\frac 12\rho_2C_{apq}*(C^{qpn}\vt_{n}-C^{qmp}\vt_{m}-C^{pqn}\vt_{n}+
C^{bqp}\vt_{b}+C^{pmq}\vt_{m}-C^{bpq}\vt_{b})\nonumber\\
&&\qquad =-\rho_2C_{apq}(C^{qpn}+C^{pnq}+C^{nqp})*\vt_{n}=
\boxed{\rho_2\Big({}^{(3)}\!A_{ab}-2 \ {}^{(4)}\!A_{ab}\Big)*\vt^b}
\er
The fourth quadratic term in equation (\ref{4-1}) using the previous one
takes the form
\br
&&\boxed{\rho_2e_a\hook\Big(d\vt^c\wedge\vt_c\wedge*(d\vt^b\wedge\vt_b)\Big)}=
\frac 12\rho_2\Big({}^{(3)}\!A_{mn}-2 \ {}^{(4)}\!A_{mn}\Big)e_a\hook(\vt^m\wedge*\vt^n)
\nonumber\\
&&\quad =\frac 12\rho_2\Big({}^{(3)}\!A_{mn}-2 \ {}^{(4)}\!A_{mn}\Big)\eta^{mn}*\vt_a
=\boxed{\frac 12\rho_2\Big({}^{(2)}\!A-2 \ {}^{(3)}\!A\Big)\eta_{ab}*\vt^b}
\er
The fifth quadratic term in equation (\ref{4-1}) is
\br
&&\boxed{-2\rho_2( e_a \hook d\vt^b) \wedge \vt_b \wedge * ( d\vt^c \wedge \vt_c)}=
-\frac 12\rho_2{C^b}_{mn}{C}_{cpq}( e_a \hook\vt^{mn})\wedge\vt_b\wedge * \vt^{pqc}\nonumber\\
&&\quad =
\frac 12\rho_2{C^b}_{mn}{C}_{cpq}( e_a \hook\vt^{mn})\wedge *(e_b\hook \vt^{pqc})\nonumber\\
&&\quad =
\frac 12\rho_2{C^b}_{mn}{C}_{cpq}(\d^m_a\vt^n-\d^n_a\vt^m)\wedge *(\d^p_n\vt^{qc}-
\d^q_n\vt^{pc}+\d^c_n\vt^{pq})\nonumber\\
&&\quad =\rho_2{{C^b}_a}^n*\Big(C_{cbn}\vt^c-C_{nbq}\vt^q-C_{cnb}\vt^c+
C_{npb}\vt^p+C_{bnq}\vt^q-C_{bpn}\vt^p\Big)\nonumber\\
&&\quad =2\rho_2{{C^m}_a}^n\Big(C_{bmn}+C_{nbm}+C_{mnb}\Big)*\vt^b=
\boxed{2\rho_2\Big({}^{(4)}\!A_{ba}+{}^{(5)}\!A_{ab}-{}^{(6)}\!A_{ab}\Big)*\vt^b}\nonumber\\
\er
The sixth quadratic term in equation (\ref{4-1}) is
\br
&&\boxed{2\rho_3 d\vt_b \wedge * ( \vt_a\wedge d \vt^b)}=
\frac12\rho_3C_{bmn}C^{bpq}\vt^{mn}\wedge *\vt_{apq}=-2C_{bmn}C^{bpq}\vt^m\wedge 
*(e^n\hook \vt_{apq})\nonumber\\
&&\quad =-\frac12\rho_3C_{bmn}C^{bpq}\vt^m\wedge *(\d^n_a\vt_{pq}-\d^n_p\vt_{aq}+\d^n_q\vt_{ap})
\nonumber\\
&&\quad =
-\frac12\rho_3C_{bmn}C^{bpq}*\Big(\d^n_a(\d^m_p\vt_q-\d^m_q\vt_p)-\d^n_p(\d^m_a\vt_q-\d^m_q\vt_a)+\d^n_q(\d^m_a\vt_p-\d^m_p\vt_a)\Big)\nonumber\\
&&\quad =-\rho_3C^{bpq}*\Big(C_{bpa}\vt_q+C_{bqp}\vt_a+C_{baq}\vt_p\Big)=
\boxed{\rho_3\Big({}^{(2)}\!A\eta_{ab}-2 \ {}^{(6)}\!A_{ab}\Big)*\vt^b}
\er
The seventh quadratic term in equation (\ref{4-1}) can be calculated using 
the relation (\ref{3-8})
\br
&&\boxed{\rho_3e_a \hook( \vt_c \wedge d\vt^b \wedge *( d\vt^c \wedge \vt_b ))}=
\rho_3e_a\hook {}^{(3)}V=\boxed{\frac12\rho_3\Big(-2 \ {}^{(1)}\!A+{}^{(2)}\!A\Big)\eta_{ab}*\vt^b}\nonumber\\
\er
The eighth quadratic term in equation (\ref{4-1}) is
\br
&&\boxed{- 2\rho_3( e_a \hook d\vt^b) \wedge \vt_c \wedge * ( d\vt^c \wedge\vt_b )}=
\frac12\rho_3{C^b}_{mn}{C_c}^{pq}\Big((e_a\hook \vt^{mn})\wedge *(e^c\hook \vt_{pqb})\Big)
\nonumber\\
&&\quad =\rho_3{C^b}_{an}{C_c}^{pq}*
\Big(e^n\hook (\d^c_p\vt_{qb}-\d^c_q\vt_{pb}+\d^c_b\vt_{pq})\Big)=
2\rho_3\Big(C_{ban}{C_m}^{mn}+C_aC_b+C_{man}{C^{mn}}_b\Big)\nonumber\\
&&\quad =
\boxed{2\rho_3\Big(-{}^{(1)}\!A_{ba}-{}^{(6)}\!A_{ab}+{}^{(7)}\!A_{ab}\Big)*\vt^b}
\er
The substitution of (\ref{2-13}-\ref{2-22}) in (\ref{3-1}) results in (\ref{3-2}).
\sect{A Diagonal Static Ansatz}
Consider a diagonal static coframe
\footnote{\rm {Greek indices run from 0 to 3 while Roman indices run from 1 to 3.}}
\begin{equation}\label{D-1}
\vt^0=e^fdx^0,\qquad \vt^m=e^gdx^m ,
\end{equation}
where $f$ and $g$ are two arbitrary functions of the spatial coordinates
$x,y,z$.
Compute the exterior derivative of the coframe
\brn
d\vt^0&=&e^ff_mdx^m\wedge dt=e^{-g}f_m\vt^{m0},\\
d\vt^k&=&e^gg_mdx^m\wedge dx^k=e^{-g}g_m\vt^{mk}.
\ern
Thus the non-vanishing $C$-objects take the form:
\br
\label{D-3}
{C^0}_{m0}&=&-{C^0}_{0m}=e^{-g}f_m,\\
\label{D-4}
{C^k}_{mn}&=&e^{-g}(g_m\d^k_n-g_n\d^k_m)
\er
or, lowering the indices
\footnote{\rm {Recall that we use the signature $(+,-,-,-)$.}}
\begin{equation}\label{D-3a}
\boxed{{C}_{0m0}=-{C}_{00m}=e^{-g}f_m}
\end{equation}
\begin{equation}\label{D-4a}
\boxed{{C}_{kmn}=-{C}_{knm}=e^{-g}(g_m\eta_{kn}-g_n\eta_{km})}
\end{equation}
The one-indexed $C$-objects $C_\a={C^\b}_{\a\b}$ are
\begin{equation}\label{D-4aa}
C_0= {C^k}_{0k}=0
\end{equation}
and
\br\label{D-4ab}
C_m&=&{C^0}_{m0}+{C^n}_{mn}=e^{-g}f_m+e^{-g}(g_m\d^n_n-g_n\d^n_m)\nonumber\\
&=&e^{-g}(f_m+2g_m)
\er
Compute the exterior derivative of the $C$-objects:
\brn
d{C^0}_{m0}&=&e^{-2g}(f_{mk}-f_mg_k)\vt^k\\
d{C^k}_{mn}&=&e^{-2g}\Big((g_{mp}-g_mg_p)\d^k_n-(g_{np}-g_ng_p)\d^k_m\Big)\vt^p.
\ern
Thus the non-vanishing four-indexed $B$-objects are 
\begin{equation}\label{D-6}
{{B^0}_{m0k}=e^{-2g}(f_{mk}-f_mg_k)},
\end{equation}
\begin{equation}\label{D-7}
{{B^k}_{mnp}=e^{-2g}\Big((g_{mp}-g_mg_p)\d^k_n-(g_{np}-g_ng_p)\d^k_m\Big)},
\end{equation}
or, lowering the first index
\begin{equation}\label{D-6a}
\boxed{{B}_{0m0k}=-{B}_{00mk}=e^{-2g}(f_{mk}-f_mg_k)}
\end{equation}
\begin{equation}\label{D-7a}
\boxed{{B}_{kmnp}=
e^{-2g}\Big((g_{mp}-g_mg_p)\eta_{kn}-(g_{np}-g_ng_p)\eta_{km}\Big)}
\end{equation}
The first two-indexed B-object is
${}^{(1)}\!B_{\a\b}=B_{\a\b\g\d}\eta^{\g\d}.$
Thus
\begin{equation}
{}^{(1)}\!B_{00}=B_{00mk}\eta^{mk}=-e^{-2g}(f_{mk}-f_mg_k)\eta^{mk}=
e^{-2g}(\triangle f-\nabla f \ \nabla g).
\end{equation}
\begin{equation}
{}^{(1)}\!B_{0m}={}^{(1)}\!B_{k0}=0
\end{equation}
\br
{}^{(1)}\!B_{km}&=&B_{kmnp}\eta^{np}=
e^{-2g}\Big((g_{mp}-g_mg_p)\eta_{kn}-(g_{np}-g_ng_p)\eta_{km}\Big)\eta^{np}
\nonumber\\
&=&e^{-2g}\Big[\eta_{km}(\triangle g-\nabla g \ \nabla g)+(g_{mk}-g_mg_k)\Big].
\er
We use here and later the following notations
\br
\triangle f&=&f_{11}+f_{22}+f_{33}=-\eta^{mn}f_{mn},\\
\nabla f \ \nabla g&=&f_1g_1+f_2g_2+f_3g_3=-\eta^{mn}f_mg_n.
\er
The second two-indexed $B$-objects  ${}^{(2)}\!B_{\a\b}=B_{\g\d\a\b}\eta^{\g\d}$ are  calculated to be 
\begin{equation}
{}^{(2)}\!B_{00}={}^{(2)}\!B_{0n}={}^{(2)}\!B_{n0}=0,
\end{equation}
\br
{}^{(2)}\!B_{np}&=&B_{00np}+B_{kmnp}\eta^{km}=
-e^{-2g}(f_{np}-f_ng_p)+\nonumber\\&&
e^{-2g}\Big((g_{mp}-g_mg_p)\eta_{kn}-(g_{np}-g_ng_p)\eta_{km}\Big)\eta^{km}
\nonumber\\&=&
e^{-2g}\Big((g_{np}-g_ng_p)-3(g_{np}-g_ng_p)-(f_{np}-f_ng_p)\Big)
\nonumber\\&=&
-e^{-2g}\Big(2(g_{np}-g_ng_p)+(f_{np}-f_ng_p)\Big)
\er
The antisymmetric $B$-object ${}^{(3)}\!B_{\a\b}=B_{\g\a\b\d}\eta^{\g\d}$ has the components
\begin{equation}
{}^{(3)}\!B_{0m}={}^{(3)}\!B_{m0}=0,
\end{equation}
\br
{}^{(3)}\!B_{mn}&=&B_{kmnp}\eta^{kp}=e^{-2g}\Big((g_{mp}-g_mg_p)\eta_{kn}-
(g_{np}-g_ng_p)\eta_{km}\Big)\eta^{kp}\nonumber\\
&=&e^{-2g}\Big((g_{mn}-g_mg_n)-(g_{nm}-g_ng_m)\Big)=0.
\er
The full contraction of the quantities ${B^a}_{nmp}$ gives
\br
B&=&{}^{(1)}\!B_{00}+{}^{(1)}\!B_{km}\eta^{km}=
e^{-2g}(\triangle f-\nabla f \ \nabla g)+
\nonumber\\&&
e^{-2g}\Big[\eta_{km}(\triangle g-\nabla g \ \nabla g)
+(g_{mk}-g_mg_k)\Big]\eta^{km}\nonumber\\&=&
e^{-2g}\Big[2(\triangle g-\nabla g \ \nabla g)
+(\triangle f-\nabla f \ \nabla g)\Big].
\er
Calculate  the first two-indexed $A$-object ${}^{(1)}\!A_{\a\b}=C_{\a\b\mu}C^\mu$
\br
{}^{(1)}\!A_{00}&=&C_{00m}C^m=e^{-2g}\eta^{mn}f_m(f_n+2g_n)\nonumber\\
&=&e^{-2g}(\nabla^2f+2\nabla f \ \nabla g).
\er
\begin{equation}
{}^{(1)}\!A_{0m}={}^{(1)}\!A_{m0}=0.
\end{equation}
\br
{}^{(1)}\!A_{mn}&=&C_{mn0}C^0+C_{mnk}C^k=
e^{-2g}(g_n\eta_{mk}-g_k\eta_{mn})(f_p+2g_p)\eta^{kp}\nonumber\\&=&
e^{-2g}\Big(g_nf_m+2g_ng_m+\eta_{mn}(2\nabla^2g+\nabla f \ \nabla g)\Big).
\er
For the second antisymmetric two-indexed $A$-object 
${}^{(2)}\!A_{\a\b}=C_{\mu\a\b}C^\mu$:
\begin{equation}
{}^{(2)}\!A_{00}=C_{\mu00}C^\mu=0,
\end{equation}
\begin{equation}
{}^{(2)}\!A_{0m}={}^{(2)}\!A_{m0}=0,
\end{equation}
\br
{}^{(2)}\!A_{mn}&=&C_{0mn}C^0+C_{kmn}C^k=e^{-2g}(g_m\eta_{kn}-g_n\eta_{km})
(f_p+2g_p)\eta^{kp}\nonumber\\
&=&e^{-2g}\Big(g_m(f_n+2g_n)-g_n(f_m+2g_m)\Big)\nonumber\\
&=&e^{-2g}(g_mf_n-g_nf_m).
\er
For the third symmetric two-indexed $A$-object 
${}^{(3)}\!A_{\a\b}=C_{\a\mu\nu}{C_b}^{\mu\nu}$:
\br
{}^{(3)}\!A_{00}&=&C_{00m}{C_0}^{0m}+C_{0m0}{C_0}^{m0}+C_{0mn}{C_0}^{mn}
\nonumber\\ 
&=&2e^{-2g}f_mf_n\eta^{mn}=-2e^{-2g}\nabla^2f,
\er
\begin{equation}
{}^{(3)}\!A_{0m}=0,
\end{equation}
\br
{}^{(3)}\!A_{mn}&=&C_{m00}{C_n}^{00}+2C_{m0k}{C_n}^{0k}+C_{mpq}{C_n}^{pq}
\nonumber\\ &=&
e^{-2g}(g_p\eta_{mq}-g_q\eta_{mp})(g_r\eta_{ns}-g_s\eta_{nr})\eta^{pr}\eta^{qs}
\nonumber\\ &=&
-2e^{-2g}(\eta_{mn}\nabla^2g+g_mg_n).
\er
For a fourth two-indexed $A$-object ${}^{(4)}\!A_{\a\b}=C_{\a\mu\nu}{{C^\mu}_b}^\nu$:
\br
{}^{(4)}\!A_{00}&=&C_{00m}{{C^0}_0}^m+C_{0m0}{{C^m}_0}^0+
C_{0mn}{{C^m}_0}^n\nonumber\\ 
&=&C_{00m}C^{00m}+C_{0m0}C^{m00}+C_{0mn}C_{m0n}=
e^{-2g}f_mf_n\eta^{mn}\nonumber\\ &=&
-e^{-2g}\nabla^2f,
\er
\begin{equation}
{}^{(4)}\!A_{0a}={}^{(4)}\!A_{a0}=0,
\end{equation}
\br
{}^{(4)}\!A_{ab}&=&C_{a0n}{{C^0}_b}^n+C_{am0}{{C^m}_b}^0+
C_{amn}{{C^m}_b}^n\nonumber\\ &=&
C_{a0n}C_{0bm}\eta^{mn}+C_{am0}C_{nb0}\eta^{mn}+C_{amn}C_{pbq}\eta^{pm}\eta^{qn}\nonumber\\ &=&
e^{-2g}(g_m\eta_{an}-g_n\eta_{am})(g_b\eta_{pq}-g_q\eta_{pb})\eta^{pm}\eta^{qn}
\nonumber\\ &=&
-e^{-2g}(\eta_{ab}\nabla^2g+g_ag_b),
\er
The fifth symmetric two-indexed $A$-object 
${}^{(5)}\!A_{\a\b}=C_{\mu\a\nu}{{C^\nu}_\b}^\mu $ is
\br
{}^{(5)}\!A_{00}&:=&C_{00n}{{C^n}_0}^0+C_{m0n}{{C^n}_0}^m=0,
\er
\begin{equation}
{}^{(5)}\!A_{0a}=0,
\end{equation}
\br
{}^{(5)}\!A_{ab}&=&C_{0 a0}{{C^0}_b}^0+
C_{0a n}{{C^n}_b}^0+C_{ma0}{{C^0}_b}^m+
C_{man}{{C^n}_b}^m\nonumber\\ 
&=&C_{0 a0}C_{0b0}+C_{0a n}C_{mb0}\eta^{mn}+
C_{ma0}C_{0bn}\eta^{mn}+
C_{man}C_{pbq}\eta^{pn}\eta^{qm}\nonumber\\ 
&=&e^{-g}f_ae^{-g}f_b+e^{-g}(g_a\eta_{mn}-g_n\eta_{ma})e^{-g}(g_b\eta_{pq}-g_q\eta_{pb})\eta^{pn}\eta^{qm}\nonumber\\&=&
e^{-2g}(f_af_b+2g_ag_b).
\er
The symmetric $A$-object ${}^{(6)}\!A_{\a\b}=C_{\mu\a\nu}{{C^\mu}_\b}^\nu$ has the 
components
\br
{}^{(6)}\!A_{00}&=&C_{00n}{{C^0}_0}^n+
C_{m0n}{{C^m}_0}^n=
C_{00n}C_{00m}\eta^{mn}+C_{m0n}C_{p0q}\eta^{pm}\eta^{qn}\nonumber\\&=&e^{-2g}f_mf_n\eta^{mn}= -e^{-2g}\nabla^2f
\er
\br
{}^{(6)}\!A_{0a}&=&C_{00n}{{C^0}_a}^n+C_{m0n}{{C^m}_a}^n=0
\er
\br
{}^{(6)}\!A_{ab}&=&C_{0 a0}{{C^0}_b}^0+C_{m a0}{{C^m}_b}^0+
C_{0 an}{{C^0}_b}^n+C_{man}{{C^m}_b}^n\nonumber\\&=&
C_{0 a0}C_{0 b0}+C_{m a0}C_{n b0}\eta^{mn}+C_{0 an}C_{0 bm}\eta^{mn}+
C_{man}C_{pbq}\eta^{pm}\eta^{qn}\nonumber\\&=&
e^{-2g}f_af_b+
e^{-2g}(g_a\eta_{mn}-g_n\eta_{ma})(g_b\eta_{pq}-g_q\eta_{pb})\eta^{pm}\eta^{qn}
\nonumber\\&=&
e^{-2g}(f_af_b+g_ag_b-\eta_{ab}\nabla^2g)
\er
For the symmetric object  ${}^{(7)}\!A_{\a\b}=C_\a C_\b$
\begin{equation}
{}^{(7)}\!A_{00}={}^{(7)}\!A_{0a}=0
\end{equation}
\br
{}^{(7)}\!A_{ab}&=&e^{-2g}(f_a+2g_a)(f_b+2g_b)\nonumber\\&=&
e^{-2g}\Big[f_af_b+2(f_ag_b+f_bg_a)+4g_ag_b\Big]
\er
The traces of the $A$-matrices are
\br
{}^{(1)}\!A&=& {}^{(1)}{A_\a}^\a={C^\a}_{\a\mu}C^\mu=-{C^\a}C_\a \nonumber\\
         &=&-e^{-2g}\eta^{mn}(f_m+2g_n)(f_n+2g_n)\nonumber\\
         &=&e^{-2g}(\nabla^2f+4\nabla f \ \nabla g+4\nabla^2g)\\
{}^{(2)}\!A&=& {}^{(3)}{A_\a}^\a=C_{\a\mu\nu}C^{\a\mu\nu}=
-2e^{-2g}(\nabla^2f+2\nabla^2g)\\
{}^{(3)}\!A&=& {}^{(4)}{A_\a}^\a=C_{\a\mu\nu}C^{\mu\a\nu}=
-e^{-2g}(\nabla^2f+2\nabla^2g)
\er
The  leading (second order) part of the equation (\ref{2-26}) 
is a linear combination of two-indexed  $B$-objects:
\br
L_{\a\b}&=&\b_1{}^{(1)}\!B_{(\a\b)}+\b_2{}^{(2)}\!B_{(\a\b)}+\b_3{}^{(3)}\!B_{\a\b}+
\b_4\eta_{\a\b}B+\nonumber\\
&&\b_5{}^{(1)}\!B_{[\a\b]}+\b_6{}^{(2)}\!B_{[\a\b]}.
\label{comb1*}
\er
The general quadratic part of the equation (\ref{2-26}) is
\br
\label{comb2*}
Q_{\a\b}&=&\a_1{}^{(1)}\!A_{(\a\b)}+\a_2{}^{(2)}\!A_{\a\b}+\a_3{}^{(3)}\!A_{\a\b}+
\a_4{}^{(4)}\!A_{(\a\b)}+\a_5{}^{(5)}\!A_{\a\b}+\nonumber \\
&&\a_6{}^{(6)}\!A_{\a\b}+\a_7{}^{(7)}\!A_{\a\b}+\a_8{}^{(1)}\!A_{[ab]}+
\a_9{}^{(4)}\!A_{[ab]}+\nonumber \\
&&
\eta_{\a\b}\Big(\a_{10}{}^{(1)}\!A+\a_{11}{}^{(2)}\!A+\a_{12}{}^{(3)}\!A\Big)
\er
Using the calculations above the leading part is  
\br
L_{00}\eqq\b_1e^{-2g}(\triangle f-\nabla f \ \nabla g)+
\b_4e^{-2g}\Big[2(\triangle g-\nabla g \ \nabla g)
+(\triangle f-\nabla f \ \nabla g)\Big]\nonumber\\
\eqq
e^{-2g}\Big[(\b_1+\b_4)\triangle f+2\b_4\triangle g
-(\b_1+\b_4)\nabla f \ \nabla g-
2\b_4\nabla g  \  \nabla g\Big]
\er
\begin{equation}
L_{0m}=L_{m0}=0
\end{equation}
\br
L_{mn}\eqq\b_1e^{-2g}\Big[
\eta_{mn}(\triangle g-\nabla g  \  \nabla g)+
(g_{mn}-g_mg_n)\Big]-\nonumber\\&&
\b_2e^{-2g}\Big(2(g_{mn}-g_mg_n)+
(f_{mn}-f_mg_n)\Big)+\nonumber\\&&
\b_4\eta_{mn}e^{-2g}\Big[2(\triangle g-\nabla g  \  \nabla g)
+(\triangle f-\nabla f \ \nabla g)\Big]
\nonumber\\\eqq 
e^{-2g}\Big[(\b_1-2\b_2)g_{mn}+(\b_1+2\b_4)\eta_{mn}\triangle g-\b_2f_{mn}+\b_4\eta_{mn}\triangle f
]-\nonumber\\&&
(\b_1-\b_2)g_mg_n
+\b_2f_mg_n
-(\b_1+2\b_4)\eta_{mn}\nabla g  \  \nabla g
-\b_4\eta_{mn}\nabla f \ \nabla g\Big]\nonumber\\&&
\er
The quadratic part of the equation takes the form
\br
Q_{00}&=&e^{-2g}\Big[
\a_1(\nabla^2f+2\nabla f \ \nabla g)
-2\a_3\nabla^2f
-\a_4\nabla^2f
-\a_6\nabla^2f\nonumber\\&&\qquad \qquad 
+\a_{10}(\nabla^2f+4\nabla f \ \nabla g+4\nabla^2g)
-2\a_{11}(\nabla^2f+2\nabla^2g)\nonumber\\&&
\qquad \qquad 
-\a_{12}(\nabla^2f+2\nabla^2g)
\Big]\nonumber\\&=&
e^{-2g}\Big[(\a_1-2\a_3-\a_4-\a_6+\a_{10}-2\a_{11}-\a_{12})\nabla^2f+\nonumber\\&&
(4\a_{10}-4\a_{11}-2\a_{12})\nabla^2g+
(2\a_1+4\a_{10})\nabla f \ \nabla g\Big]
\er
\begin{equation}
Q_{0m}=Q_{m0}=0
\end{equation}
\br
Q_{mn}&=&e^{-2g}\Big[
\a_1\Big(g_nf_m+2g_ng_m+\eta_{mn}(2\nabla^2g+\nabla f \ \nabla g)\Big)
\nonumber\\&&\quad\quad
+\a_2(g_mf_n-g_nf_m)
-2\a_3(\eta_{mn}\nabla^2g+g_mg_n)\nonumber\\&&\quad\quad
-\a_4(\eta_{mn}\nabla^2g+g_mg_n)
+\a_5(f_mf_n+2g_mg_n)\nonumber\\&&\quad\quad
+\a_6(f_mf_n+g_mg_n-\eta_{mn}\nabla^2g)
\nonumber\\&&\quad\quad
+\a_7\Big(f_mf_n+2(f_mg_n+g_mf_n)+4g_mg_n\Big)
\nonumber\\&&\quad\quad
+\eta_{mn}\Big(
\a_{10}(\nabla^2f+4\nabla f \ \nabla g+4\nabla^2g)\Big)
\nonumber\\&&\quad\quad
-2\a_{11}(\nabla^2f+2\nabla^2g)
-\a_{12}(\nabla^2f+2\nabla^2g)
\Big]
\nonumber\\&=&
e^{-2g}\Big[
(\a_5+\a_6+\a_7)f_mf_n+
(\a_2+\a_7)g_mf_n+(-\a_2+2\a_7)g_nf_m+\nonumber\\&&\quad\quad
(2\a_1-2\a_3-\a_4+2\a_5+\a_6+4\a_7)g_mg_n+\nonumber\\&&\quad\quad
\eta_{mn}\Big(
(\a_{10}-2\a_{11}-\a_{12})\nabla^2f+(\a_1+4\a_{10})\nabla f \ \nabla g+
\nonumber\\&&\quad\quad
(2\a_1-2\a_3-\a_4-\a_6+4\a_{10}-4\a_{11}-2\a_{12})\nabla^2g
\Big)\Big]
\er
Thus the field equation (\ref{2-26}) reduces to the form
\begin{equation}\boxed{
\mu_1\triangle f+\mu_2\triangle g=\mu_3\nabla^2f+\mu_4\nabla^2g+
\mu_5(\nabla f \ \nabla g)}
\end{equation}
\begin{equation}\boxed{
\begin{array}{ll}
&\eta_{mn}(\nu_1\triangle f+\nu_2\triangle g)+\nu_3 f_{mn}+\nu_4g_{mn}=\nonumber\\&
\qquad \qquad\eta_{mn}(\nu_5\nabla^2f+\nu_6\nabla^2g+\nu_7(\nabla f \ \nabla g))+\nonumber\\
&\qquad \qquad  \nu_8 f_mf_n+\nu_9g_mg_n+\nu_{10}f_mg_n+\nu_{11}f_ng_m,
\end{array}}
\end{equation}
where the numerical coefficients are (\ref{5-4}).

}
\end{document}